\documentstyle[aaspp4]{article}

\begin{document}

\title{Evolution of the Low-Energy Photon Spectra\\
	 in Gamma-Ray Bursts}

\author{A. Crider, E. P. Liang, and I. A. Smith}
\affil{Department of Space Physics and Astronomy, 6100 S. Main,
 	Rice University, Houston, TX 77005-1892}

\author{R. D. Preece, M. S. Briggs, G. N. Pendleton, and W. S. Paciesas}
\affil{Department of Physics, University of Alabama in Huntsville,
 	Huntsville, AL  35899}

\and

\author{D. L. Band and J. L. Matteson}
\affil{Center for Astrophysics and Space Sciences 0111, University of California
	at San Diego, La Jolla, CA 92093}

\received{15 Nov 1996}
\accepted{6 Dec 1996}

\abstract{We report evidence that the asymptotic low-energy power law slope
$\alpha$ (below the spectral break) of BATSE gamma-ray burst photon spectra
evolves with time rather than remaining constant.
We find that a high degree of positive correlation exists between 
the time-resolved spectral break energy $E_{\rm{pk}}$ and $\alpha$.
In samples of 18 ``hard-to-soft'' and 12 ``tracking'' pulses, 
evolution of $\alpha$ was found to correlate with that of the spectral break
energy $E_{\rm{pk}}$ at the 99.7\% and 98\% confidence levels respectively.
We also find that in the flux rise phase of ``hard-to-soft'' pulses,
the mean value of $\alpha$ is often positive and in some bursts 
the maximum value of $\alpha$ is consistent with a value $> +1$.
BATSE burst 3B 910927,
for example, has an $\alpha_{\rm{max}}$ equal to $1.6~\pm~0.3$. 
These findings challenge GRB spectral models
in which $\alpha$ must be negative or remain constant.}


\section{Introduction}

Studies of gamma-ray burst (GRB) spectral evolution have recently 
begun to uncover trends which may constrain the emission 
mechanisms.  Earlier reports on spectral evolution 
focused solely on the ``hardness'' of bursts, measured either by the  
ratio between two detector channels or with 
more physical variables such as the spectral break
or peak power energy $E_{\rm{pk}}$ (\cite{ford95})
which is the maximum of $\nu F_{\nu}$, where
$\nu$ is photon energy and $F_{\nu}$ is the specific energy flux.
Such hardness parameters were typically found to either follow a 
``hard-to-soft'' trend (\cite{norr86}),
decreasing monotonically while the flux rises and falls,
or to ``track'' the flux during GRB pulses (\cite{gole83}).

The recent discovery that $E_{\rm{pk}}$ often decays exponentially in bright, 
long, smooth BATSE GRB pulses 
\textit{as a function of photon fluence} $\Phi$
provides a new constraint for emission models (\cite{lian96}), 
and the fact that the decay constant
$\Phi_{\rm{0}}$ often remains fixed from pulse to pulse
within a single burst hints at a regenerative source rather than a  
single catastrophic event (\cite{mesz93}). 
However, that study 
concentrated only on the evolution of $E_{\rm{pk}}$.
To further explore the origin of the spectral break,
we begin the analysis of two additional parameters in  
the spectral evolution, the asymptotic low-energy power law slope $\alpha$
below 
$E_{\rm{pk}}$ and the high-energy power law slope $\beta$ above $E_{\rm{pk}}$
as they are defined in the Band {\it et al.} (1993) GRB spectral function 

\( \begin{array}{rclrr}
N_{\rm{E}}(E) & = & A\left(\frac{E}{100 \rm{keV}}\right)^{\alpha} 
   exp\left(-\frac{E}{E_{\rm{0}}}\right), & (\alpha-\beta)E_{\rm{0}} \geq E, \\
\mbox{} & = & A\left[\frac{(\alpha-\beta)E_{\rm{0}}}{100 \rm{keV}}\right]^{\alpha-
   \beta} exp(\beta-\alpha) \left(\frac{E}{\rm{100 keV}}\right)^{\beta},
& (\alpha-\beta)E_{\rm{0}} \leq E, & (1)
\end{array} \)
\vspace{2mm}

\noindent where A is the amplitude (in $\rm{photons} \:
\rm{sec}^{-1} \rm{cm}^{-2}
\rm{kev}^{-1}$) and $E_{0} = E_{\rm{pk}} / (2 + \alpha)$. 
We note that $\alpha$ is not the maximum low energy slope of the 
GRB function within the detector range, but is the asymptotic limit of the 
slope if extrapolated to arbitrarily low energies.
The observed values and variability of all three parameters
are crucial in evaluating the wide field of proposed models of
gamma-ray burst emission.  For example, many models of the spectral break 
require $\alpha$ to stay constant (e.g. self-absorption) or to have negative
values (e.g. $-\frac{2}{3}$, \cite{katz94}, \cite{tava96}), even when
$E_{\rm{pk}}$ evolves.  In our study, 
we find that there are hints that $\beta$ decreases over the course of some
bursts, as suggested by COMPTEL results (\cite{hanl95})
and stays constant in others.
In this letter, though, we focus on the evolution of $\alpha$
and save the discussion of $\beta$ for future work (\cite{pree97}). 

\section{Spectral Evolution Patterns}

To determine the evolution of the spectral shape of GRBs,
we examine High Energy Resolution
data collected from the BATSE Large-Area Detectors (LADs) on board the Compton
Gamma-Ray Observatory (\cite{fish89}). 
We select bursts that have either a BATSE 3B catalog
(\cite{meeg96}) fluence $> 2 \times 10^{-5}$ erg cm$^{-2}$, peak photon fluxes
$\stackrel{>}{\sim}$ 10 photons s$^{-1}$
cm$^{-2}$ on the 1024 ms timescale, or a signal-to-noise ratio (SNR) $> 7.5$
between 25 and 35 keV with the BATSE Spectral Detector (SD).
The inclusion of a few other bursts results in a set of 79 bursts. 
The counts from the LAD most normal to the line of sight of each burst are
background-subtracted and then binned into  
time intervals each with a SNR of $\sim 45$ within the 28 keV to 1800 keV range.  
Employing a non-linear $\chi^{2}$-minimization algorithm (\cite{bevi69}),
we fit the Band {\it et al.} GRB function 
to each interval and thus obtain the time evolution of the three Band
{\it et al.} parameters which define the spectral shape.

Figures 1 through 3 show 
sample BATSE bursts (3B) 910807, 910927, 911031, 920525, and 931126,
displayed for illustrative purposes.
(See \cite{lian96} for $E_{\rm{pk}}$-fluence
diagrams for 911031, 920525, and 931126.) 
The spectra show that for these bursts, $\alpha$ generally rises and falls
with the instantaneous $E_{\rm{pk}}$, though exact correlation between
the two parameters is not evident.
In Figures 1 and 2, $\beta$ stays relatively constant throughout most
of the primary pulse, while $E_{\rm{pk}}$ and $\alpha$ both steadily decrease.

To determine if $\alpha$ does indeed evolve in time in a majority of bursts, 
we fit a zeroth (M=0) and first-order (M=1) polynomial to the $\alpha$
evolution in each burst.
Assuming a null hypothesis in which $\alpha$ is constant during a burst and
the time-resolved values of $\alpha$ are normally distributed about the mean,
we expect the value $\Delta\chi^{2}=\chi^2_{\rm{M=0}}-\chi^2_{\rm{M=1}}$
to be distributed as $\chi^2$ with 1 degree of freedom (\cite{eadi71}). 
We calculate for each burst the probability
Q of randomly drawing a value greater than or equal to $\Delta\chi^{2}$. 
We observe that 67 of the 79 bursts have a Q $\le$ 0.05, (which gives a D=0.8
in a K-S test) and 46 have a Q below our acceptable cutoff of 0.001.
We conclude that a majority of bursts in our sample show evidence for at 
least a first-order trend in $\alpha$. 

The five sample bursts above suggest that evolution of $\alpha$ mimics that of 
$E_{\rm{pk}}$.  To see if this occurs in other bursts, we
attempt to disprove the null hypothesis that $\alpha$ is uncorrelated with
$E_{\rm{pk}}$.  To test the degree 
of correlation between $\alpha$ and $E_{\rm{pk}}$, in each of our 79 bursts
we compute the Spearman rank correlation $r_{\rm{s}}$ (\cite{pres92}).
For each burst with a positive $r_{\rm{s}}$, we find the probability $P_{+}$ of 
randomly drawing a value of $r_{\rm{s}}$ that high or higher assuming no
correlation exists.  For each burst with a negative $r_{\rm{s}}$,
we find the probability $P_{\rm{-}}$ of randomly drawing a value of 
$r_{\rm{s}}$ that low or lower assuming no anti-correlation. 
The divisions of the bursts in this way precludes the inclusion of systematic
anti-correlations, which could occur given the negative covariance between
$\alpha$ and $E_{\rm{pk}}$ and the observed shape of our $\chi^{2}$ minimum
contours.
We next calculate the Kolmogorov-Smirnov (K-S) D statistic between the 
measured distribution of $P_{\rm{+}}$ or $P_{\rm{-}}$
and the distribution one would expect if no correlation
or anti-correlation existed.  We find D=0.45 for the 47 positively correlated
bursts.  The likelihood of this value, assuming no intrinsic correlation, is 
$2\times10^{-8}$.  The bursts showing negative correlation, which suffer from 
systematics described above, are still consistent (likelihood = 0.04) with
a non-correlation hypothesis.  From this we conclude that a positive correlation 
exists between $\alpha$ and $E_{\rm{pk}}$ in at least some subset of bursts.   

To determine if this relation exists in ``hard-to-soft'' or ``tracking'' pulses,
we select 18 pulses which we determine to be clearly ``hard-to-soft'' and 12 
pulses which are clearly ``tracking''
from the $>240$ pulses within our 79 bursts.
Pulses are included in the ``hard-to-soft'' category if the maximum 
$E_{\rm{pk}}$ occurs before the flux peak and is greater than $E_{\rm{pk}}$
at the flux peak by at least $\sigma_{\rm{E_{\rm{pk}}}}$.
Pulses are ``tracking'' if the rise and fall of $E_{\rm{pk}}$ coincides with
those of flux to within 1 time bin (typically $\sim \frac{1}{2}$ sec)
and if the rise lasts at least 3 time bins.
We do not pretend that
all pulses fall into one of these two categories, but instead treat them as 
extreme examples in a continuum of evolutionary patterns.  
Following the same analysis described above on these smaller populations, 
we find D=0.46 for ``hard-to-soft'' pulses and D=0.45 for ``tracking'' pulses. 
in cases with positive $r_{\rm{s}}$.
The likelihood of these observed values of $r_{\rm{s}}$
assuming no intrinsic correlations is 0.003 for the ``hard-to-soft'' pulses
and 0.02 for the ``tracking'' pulses.  In contrast, 
while 4 of the 18 ``hard-to-soft'' pulses and 
6 of the 12 ``tracking'' pulses were anti-correlated, the likelihood of these
randomly occurring was 0.78 for the ``hard-to-soft'' cases and 0.29 for the
``tracking'' cases, values consistent with the null hypothesis of no 
anti-correlation.  
In Figure~4, we compare the cumulative distributions of the 14 ``hard-to-soft''
and 6 ``tracking'' pulses which are positively correlated to that
of the 47 positively
correlated bursts.  We find that both distributions of pulses are similar to the
distribution of the bursts which implies an $E_{\rm{pk}}-\alpha$ correlation.
We conclude from this statistical evidence that for ``hard-to-soft'' and, 
with less confidence, for ``tracking'' pulses the asymptotic low-energy
power-law slope $\alpha$ evolves in a manner similar to $E_{\rm{pk}}$. 

\section{Low-Energy Power Index in the Rise Phase of Pulses}

Assuming that $\alpha$ mimics $E_{\rm{pk}}$, 
it follows that \textit{$\alpha$ decreases monotonically for ``hard-to-soft''
pulses}, whereas it increases during the rise phase of ``tracking'' pulses.  We
compare the averaged values of $\alpha$ during the rise phase for
these two groups and find that \textit{those in ``hard-to-soft'' pulses are
significantly higher}.  While none of the 12 ``tracking'' pulses has an
average $\alpha_{\rm{rise}}~>~0$,
7 of the 18 ``hard-to-soft'' pulses had an average $\alpha_{\rm{rise}} > 0$
(see Figure~5).  A K-S test between the two distributions gives a value of
D=0.56, implying a probability of 0.014 that these two samples were
randomly taken from the same distribution.

We next examine the highest value of $\alpha_{max}$ 
that occurs in our time-resolved spectra.  This value 
serves as a valuable test for GRB emission models.
In Figure~6, we provide the distribution of $\alpha_{max}$
found in each of our 79 bursts.
Only a few bursts examined so far suggest 
that their maximum $\alpha$ may be $> +1$. 
As indicated in Figure~6, all of the bursts with $\alpha_{max} > +1$
have large statistical uncertainties. The nearly linear decrease
of $\alpha$ with respect to time in 3B 910927 suggests that its 
relatively high $\alpha_{\rm{max}}$ of $1.6~\pm~0.3$, found using data from the
LAD most normal to the burst,
is not merely a statistical fluctuation (see Figure~1).
Further examination reveals, however, that this burst is still 
consistent with  $\alpha \le +1$ for its duration.
In addition, jointly fitting the data from the two LADs most normal to the 
burst reduces $\alpha_{\rm{max}}$ to $1.03~\pm~0.15$.

We also note that fitting 3B 910927 with a broken power law instead of the Band
{\it et al.} GRB function, gives \textit{the same linear decrease of the 
low-energy power law slope $\gamma_{\rm{1}}$ with respect to time}.
The fit for the broken power law also has reduced-$\chi^{2}$ values
comparable to those of the Band {\it et al.} GRB function fit.
However, $\gamma_{\rm{1}}~<~0$ throughout the burst, a value of the low-energy
slope lower than that found using the Band {\it et al.} GRB function.  
If the GRB function better represents the underlying physics than this difference
would be expected. The parameter $\gamma_{\rm{1}}$ measures the effective
average slope below $E_{\rm{pk}}$ whereas $\alpha$ measures asymptotic value,
allowing for the curvature of the exponential function.   

\section{Summary and Discussion}

We establish that the asymptotic low-energy power law slope, represented by 
the Band {\it et al.} parameter $\alpha$, evolves 
with time rather than remaining fixed to its time-integrated value in 58\%
of the bursts in our sample.  
We find strong evidence that a correlation between the parameters $E_{\rm{pk}}$
and $\alpha$ exists in the time-resolved spectra of some BATSE gamma-ray bursts
and with slightly less confidence, we determine that this correlation exists in
both ``hard-to-soft'' and ``tracking'' pulses.
We also find that in $\sim 40\%$ of the ``hard-to-soft'' pulses,
the average value of $\alpha$
during the flux rise phase is $> 0$, while for ``tracking'' pulses
the average $\alpha$ is always $\le 0$.
For 3B 910927, using data from only the LAD receiving the most counts,
we determine a maximum value of $\alpha = 1.6\pm0.3$.
However, we cannot yet prove that $\alpha_{\rm{max}}~>~1$ in any burst examined
so far due to broadness of the $\chi^{2}$ minimum.

GRB spectral breaks can in principle be caused by
synchrotron emission with a low-energy cutoff or self-absorption.  However, in
the former case, $\alpha$ is always $\leq -\frac{2}{3}$ (\cite{katz94}, 
\cite{tava96}) with no evolution.  Such a low and constant 
$\alpha$ is inconsistent with many  observed BATSE bursts.  For instance, in
3B 910927, fitting the time bin in which $\alpha$
is maximum with an $\alpha$ fixed to $-\frac{2}{3}$ results in a Q
of $1.5 \times 10^{-11}$, much lower the Q=0.35 obtained
when $\alpha$ is a free parameter.
In the case of self-absorption, $\alpha$ could go as high
as +1 (thermal) or +1.5 (nonthermal, power-law) (\cite{rybi79}). But again, 
in such models, $\alpha$ cannot
evolve with time, only $E_{\rm{pk}}$ (which would be interpreted as the 
self-absorption frequency) can.
Hence, these conventional interpretations of the spectral break of GRB
continua can be ruled out by our results.  Implications of our results on
various cosmological scenarios (e.g. \cite{shav96}) remain to be be
investigated. 

The spectral breaks can also be caused by multiple Compton scattering
(\cite{lian96}). In this case,
the decay of $\alpha$ in ``hard-to-soft'' pulses can be 
interpreted as the Thomson thinning of a Comptonizing plasma 
(\cite{lian97}) and the initial $\alpha$ can in principle go as high as +2,
because in the limit $\tau_{\rm{T}}$ (Thomson depth) $\rightarrow \infty$ one
would expect a Wien peak. 
However, several factors make it difficult to 
clearly measure an early low-energy power law $\sim$~+2 even if the spectral
break is related to a Wien peak. 
The most obvious problem is that the
highest $\tau_{\rm{T}}$ would occur earliest in a ``hard-to-soft'' pulse,
when the flux is the lowest, so that fitting a precise spectral model
becomes difficult.  
Another problem is that even if the true GRB spectral break is Wien-like, 
if one had used a function other than the Band {\it et al.} function
(eg. broken power law) or simply measured the slope
within the BATSE range, one could get a slope flatter than +2. 
This is evident in Figure~6, in which the maximum slope for the same set of
bursts appears to only approach +1 while $\alpha_{\rm{max}}$ appears to 
approach +2.  This is
because the exponential curvature depresses the apparent slope relative to
the asymptotic power law of the Wien function.
Also important is that the Band {\it et al.} GRB function
does not take into account the soft X-ray upturn expected from saturated
Comptonization of soft photons (\cite{rybi79}, \cite{lian97}). 
If the lower boundary of the fitting energy window 
is below the relative minimum in the saturated Comptonization photon spectrum
(\cite{pozd83}),
any fitted, low-energy power law, such as the
Band {\it et al.} GRB function, will be flatter than the true slope 
for the Wien peak.
Preliminary results show that moving the lower energy cutoff of the fitting
region allows one to get a higher $\alpha$. 
However, the uncertainty in $\alpha$
increases when reducing the size of the fitting window and thus the higher
value of $\alpha$
may be misleading. Evidence for the X-ray upturns in the low-energy
spectra have been found by Preece {\it et al.} (1996) who found 
positive residuals between the BATSE data and 
their fitted Band {\it et al.} GRB functions in many bursts.
However, further analysis of time-resolved, low-energy GRB spectra is
needed before a model involving saturated Comptonization can be tested.

\acknowledgements

AC thanks NASA-MSFC for the Graduate Student Research Program
fellowship.  This work is supported by NASA grant NAG5-1515.


\figcaption [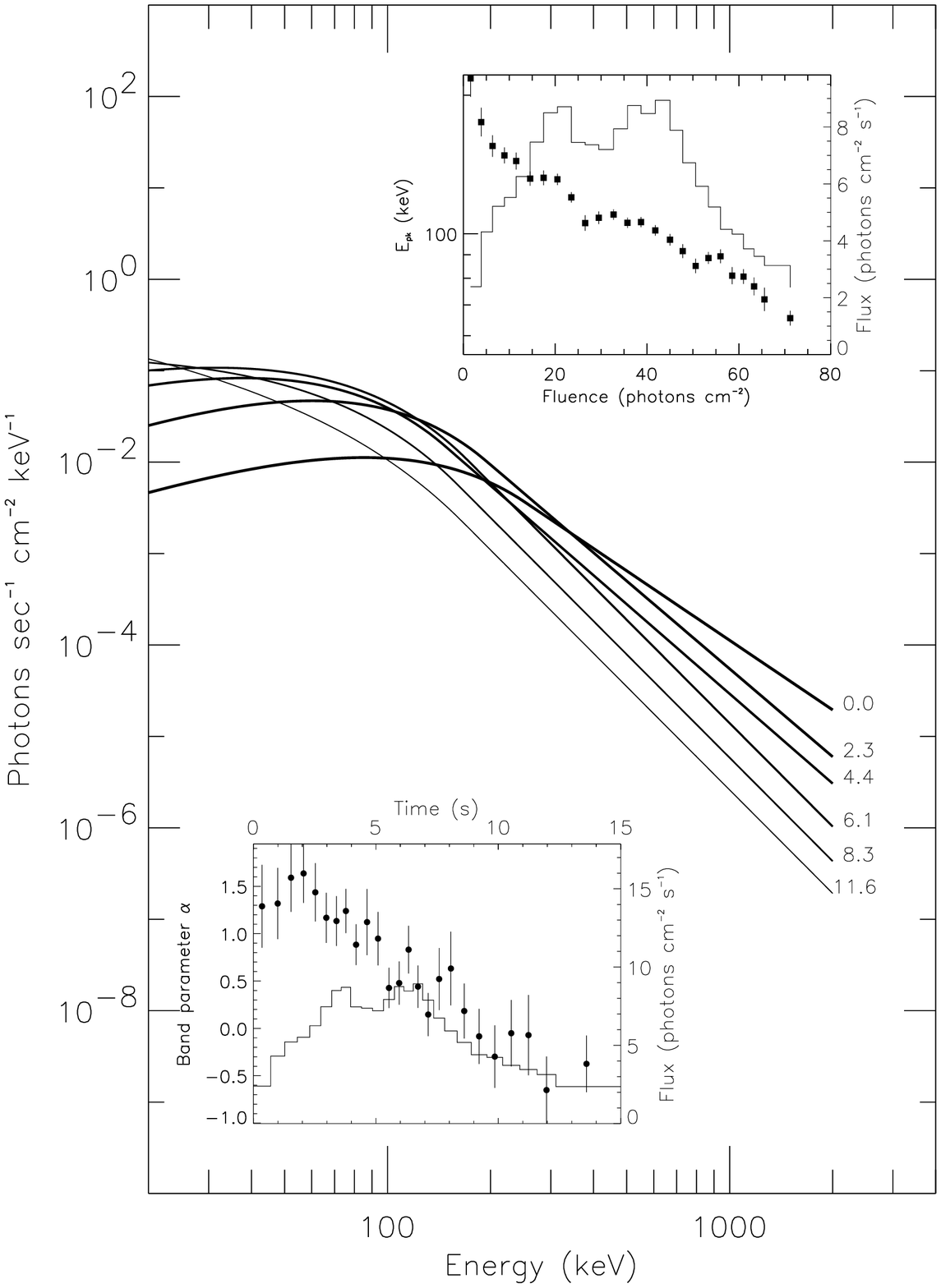] {Evolution of the Band {\it et al.} GRB spectral 
function for 3B 910927.  Each line is marked with the
time (in s) corresponding to the beginning of the time bin. 
Note that a typical statistical $\beta$ error $\sigma_{\beta} \approx 0.4$.
The nearly linear
decay of $\alpha$ throughout this burst suggests that the early high values of
$\alpha$ are not statistical fluctuations.  However, this burst is still 
consistent with $\alpha \le +1$. 
[Upper inset]
Evolution of $E_{\rm{pk}}$ (squares, logarithmic scale) and photon flux
(histogram, linear scale) with respect to fluence.  
[Lower inset] Evolution of $\alpha$ (circles) and photon flux
(histogram) with respect to time.  Error bars represent 1$\sigma$ confidence
level.}

\figcaption [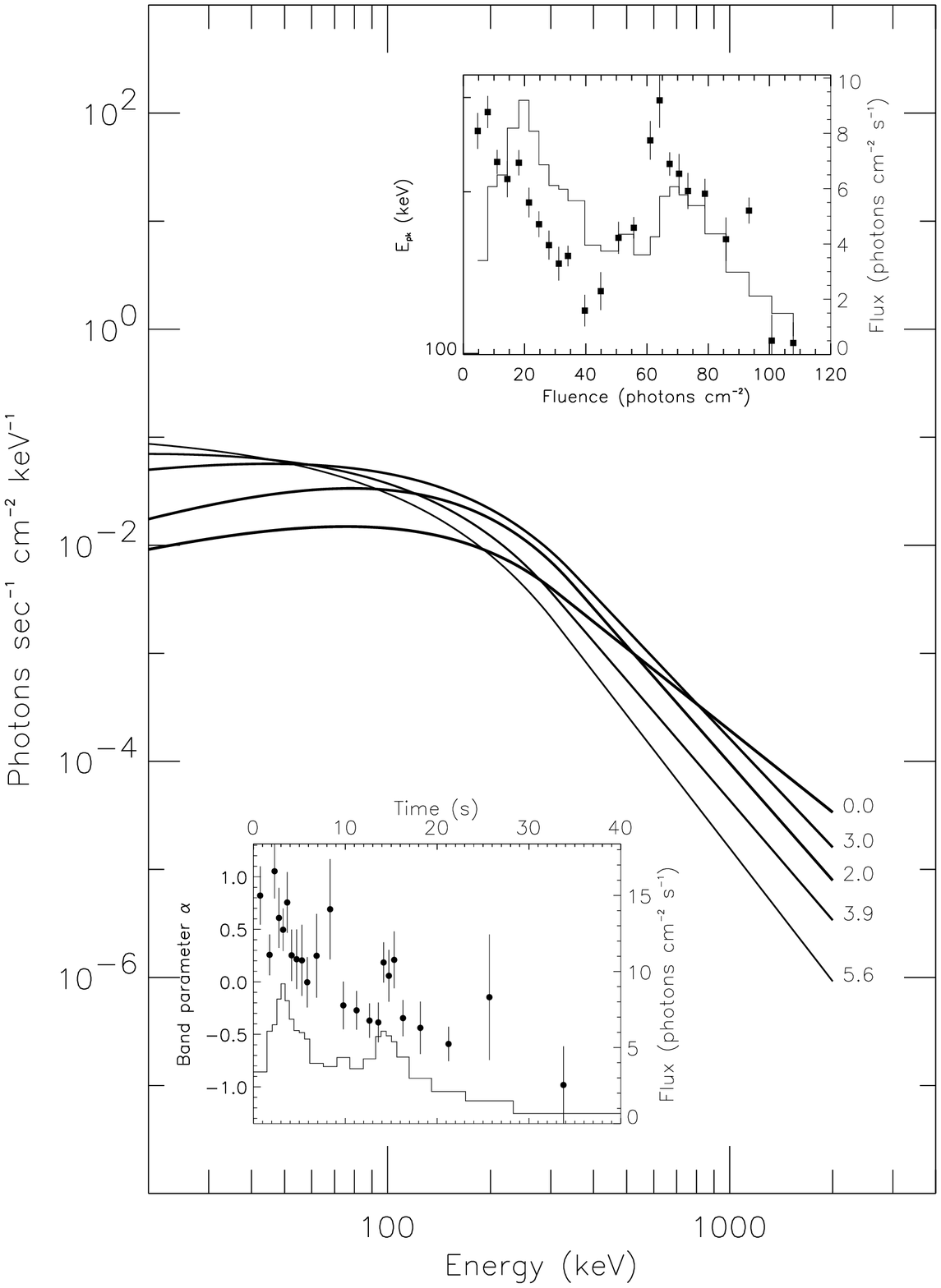] {Evolution of the Band {\it et al.} GRB spectral
function of the first pulse in 3B 910807. Note that the typical statistical
$\sigma_{\beta} \approx 0.4$.  See Figure 1 caption for description of plots.}

\figcaption [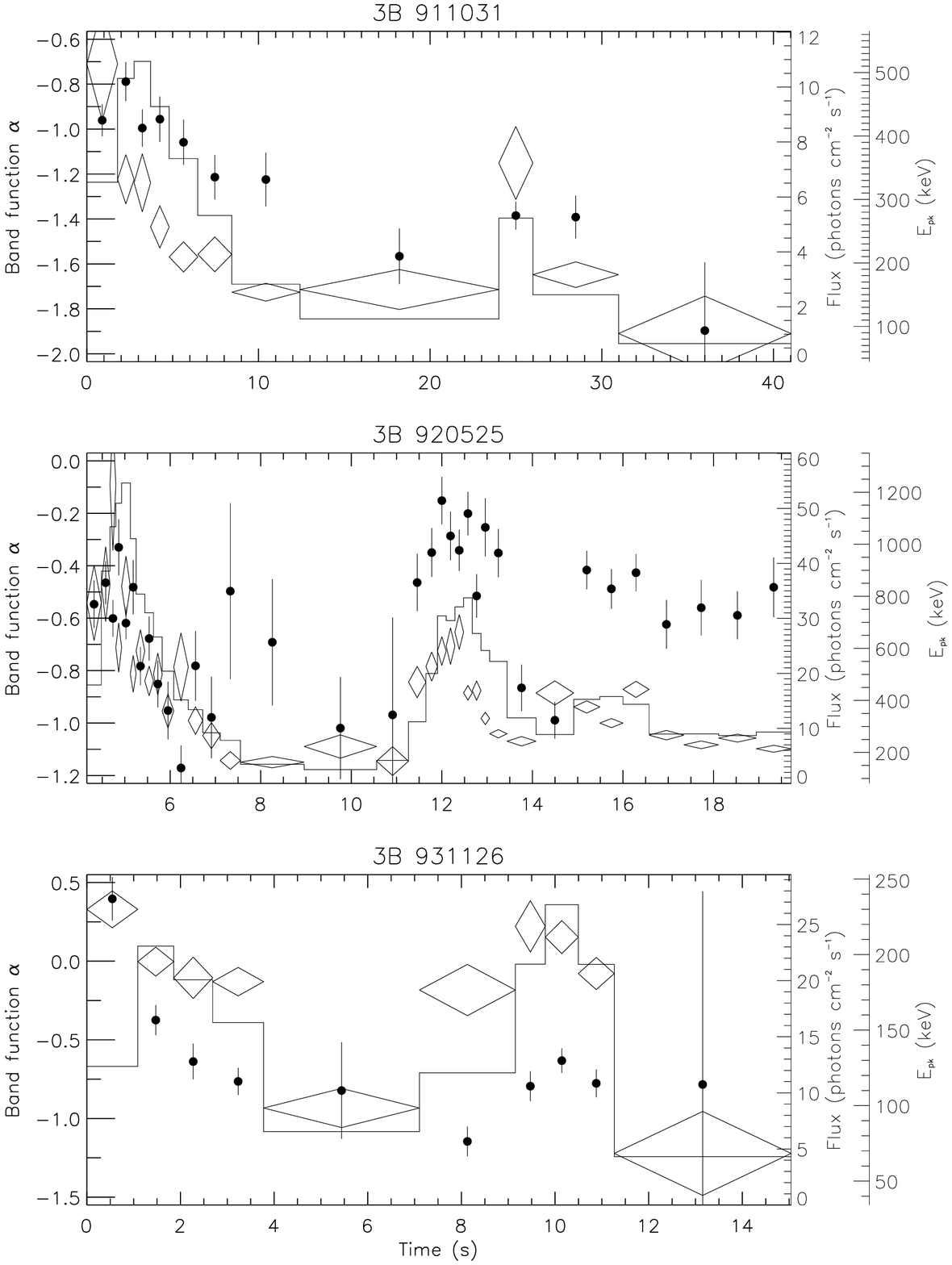] {Band {\it et al.} $\alpha$ (circles), linear 
$E_{\rm{pk}}$ (diamonds) and photon flux 
(histogram) evolution in time for 3B 911031 [GRB 973], 3B 920525 [GRB 1625],
and 3B 931126 [GRB 2661]. These three bursts suggest that $\alpha$ evolves in
a manner similar to $E_{\rm{pk}}$.  Error bars represent 1$\sigma$ confidence region.
The vertical dimensions of diamonds represents 1$\sigma$ confidence
region for $E_{\rm{pk}}$ and the horizontal dimensions represent
the durations of the time bins. 
(See Liang \& Kargatis 1996 for $E_{\rm{pk}}$ vs.
fluence of these bursts with $\alpha$ and $\beta$ fixed 
to the time-integrated value.)}

\figcaption [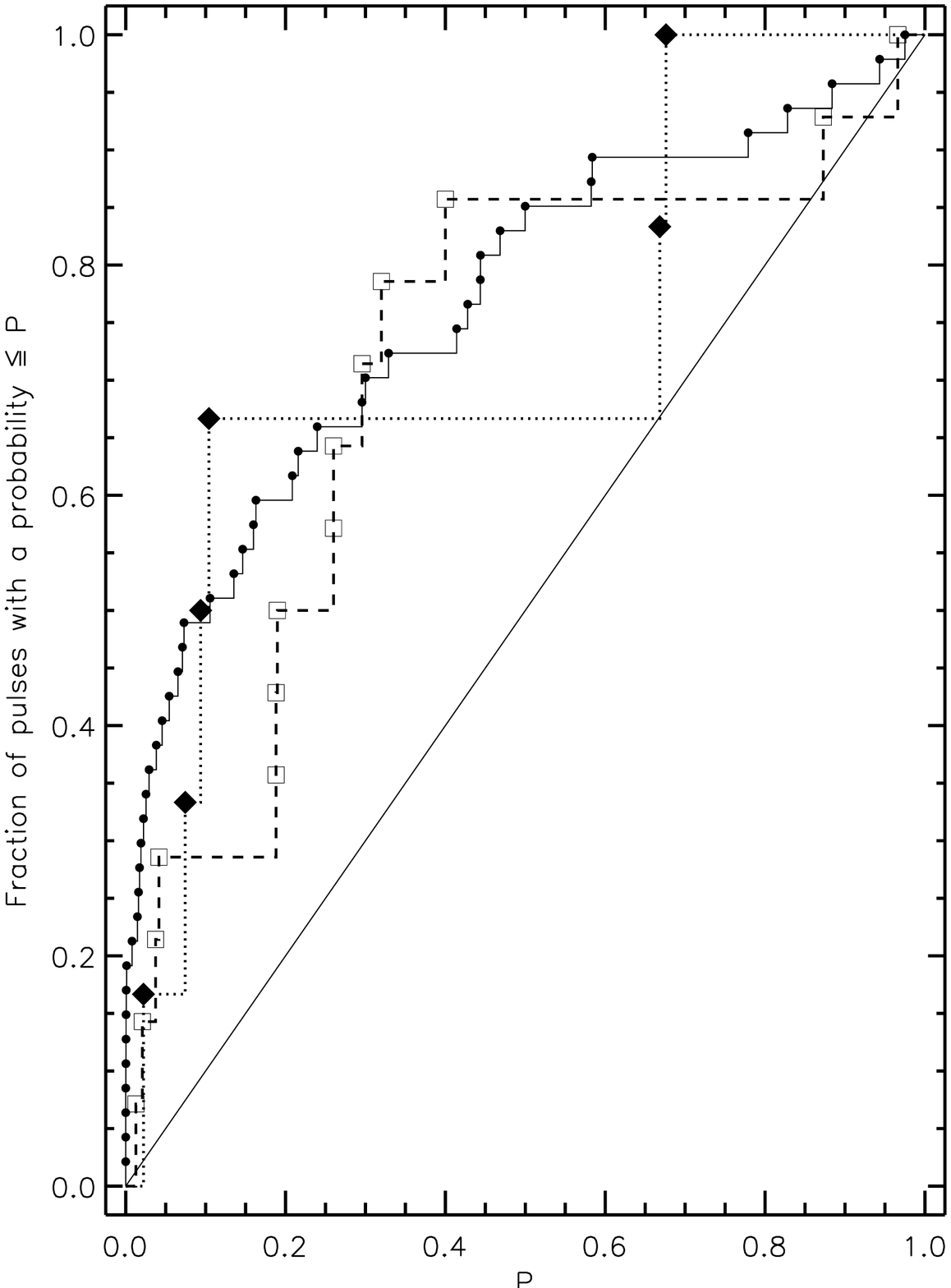] {Cumulative distribution of 
the probability $P$ of randomly drawing a value $\geq t$
assuming no correlation between $E_{\rm{pk}}$ and $\alpha$
for 14 ``hard-to-soft'' pulses
(squares), 6 ``tracking'' pulses (diamonds),
and 47 bursts (circles) with positive
correlation.  Applying a K-S test, we find the probabilities of getting
these distributions are, respectively, 0.003, 0.02, and $2 \times 10^{-8}$.  
The probabilities (from a K-S test) that the positively-correlated 
``hard-to-soft'' and ``tracking'' pulses represent samples from the same
population from which the 47 bursts are taken are, respectively, 0.21 and 0.55.}

\figcaption [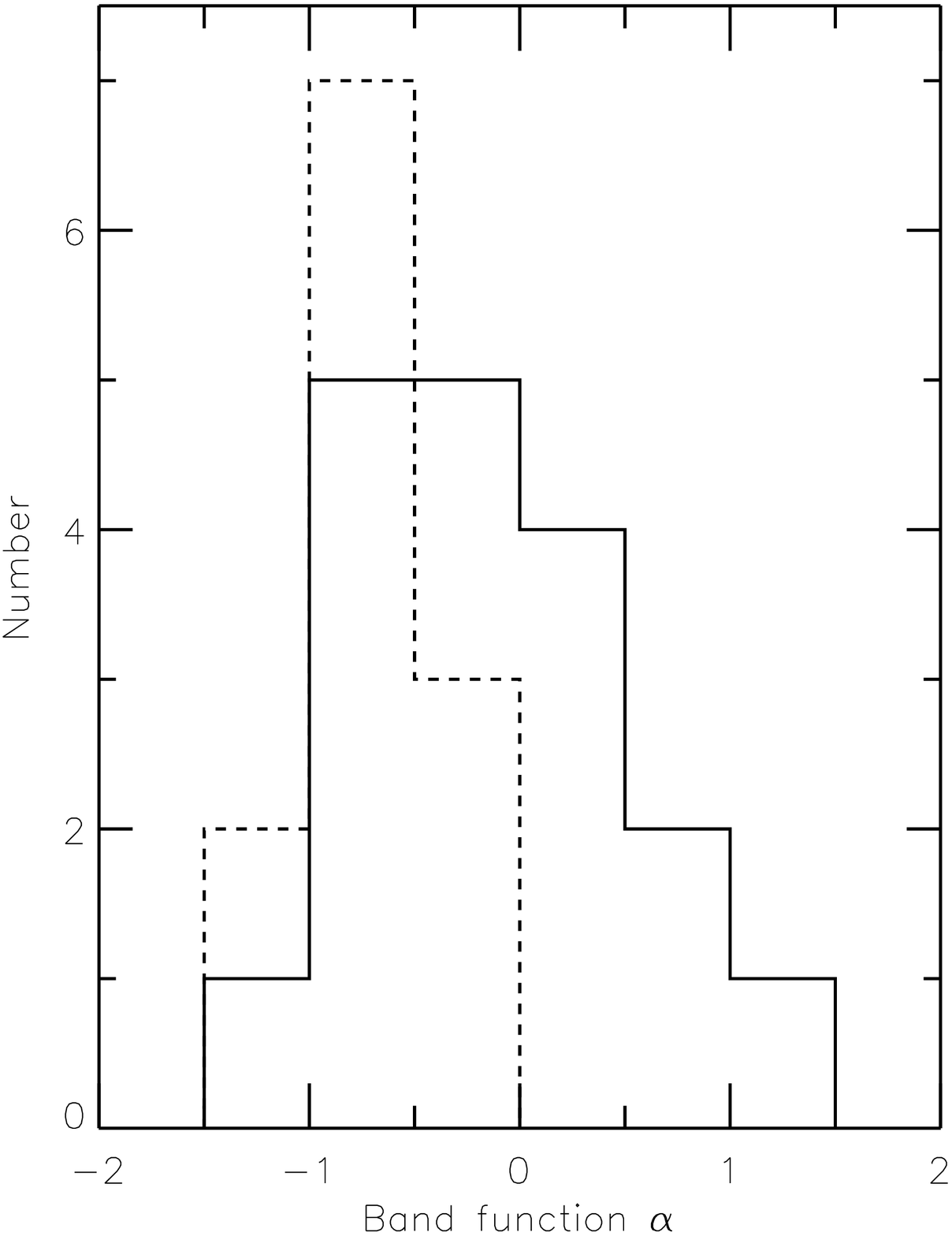] {Mean value of Band {\it et al.} parameter $\alpha$
for the rise phase of 18 ``hard-to-soft'' pulses (solid) and 12 ``tracking''
(dashed) pulses.  Note that only ``hard-to-soft'' pulses have mean values of
$\alpha_{\rm{rise}} > 0$.  The probability of these two samples originating from
the same population is 0.014 as determined by the K-S test.}

\figcaption [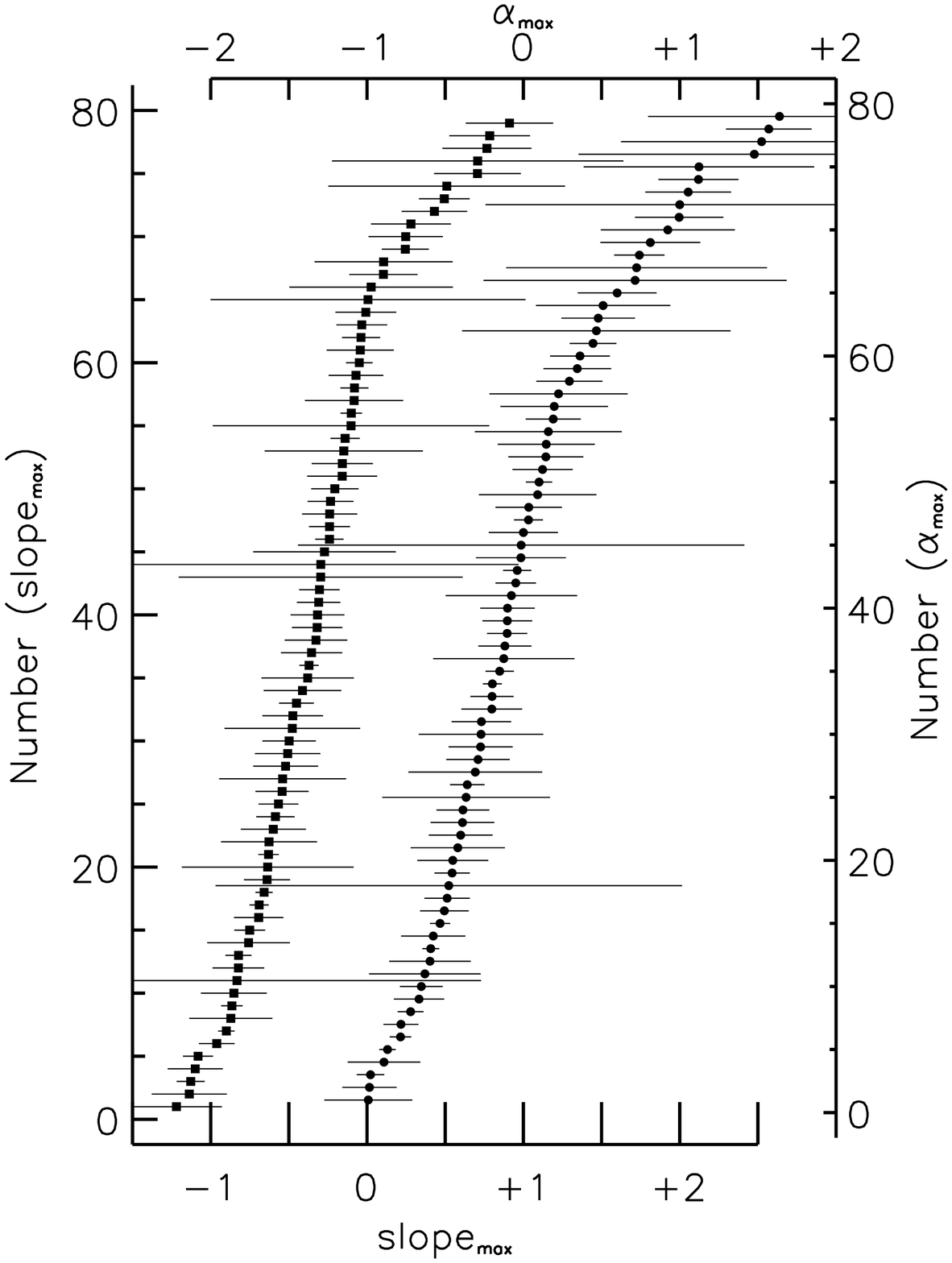] {Cumulative distribution of the maximum value of 
the Band {\it et al.} parameter $\alpha$ (circles) and 
the maximum spectral slope (squares) within the fitted energy region for  
79 BATSE bursts.  The slope (in log-log space) of the Band {\it et al.}
GRB function equals $\alpha - E/E_{\rm{0}}$ when $E \leq 
(\alpha - \beta) E_{\rm{0}}$ and equals $\beta$ when $E \geq (\alpha - \beta)
E_{\rm{0}}$, where $E_{\rm{0}} = E_{\rm{pk}}/(2 + \alpha)$.    
Error bars represent the 1$\sigma$ confidence region.
Note that $\rm{slope}_{\rm{max}}$ represents
the maximum slope measured in each burst and does {\it not} necessarily
occur at the time when the Band {\it et al.} function $\alpha$
is maximum.  Also note that these are two separate distributions
and the \textit{i}th $\alpha_{\rm{max}}$ is not necessarily from the same burst
as the \textit{i}th $\rm{slope}_{\rm{max}}$. }

\clearpage
\plotone{Figure1.eps}
\clearpage
\plotone{Figure2.eps}
\clearpage
\plotone{Figure3.eps}
\clearpage
\plotone{Figure4.eps}
\clearpage
\plotone{Figure5.eps}
\clearpage
\plotone{Figure6.eps}

\end{document}